\documentclass[prb,preprint,showpacs]{revtex4}
\usepackage{epsfig}

\begin{document}
\title{Exact one- and two-particle excitation spectra of acute-angle
helimagnets above their saturation magnetic field}
\author{R.O. Kuzian}
\affiliation{Institute for Problems of Materials Science Krzhizhanovskogo 3,\\
03180 Kiev, Ukraine}
\author{S.-L. Drechsler}
\affiliation{Leibniz-Institut f\"ur Festk\"orper- und
Werkstofforschung IFW Dresden, P.O. Box 270116, D-01171 Dresden,
Germany}

\begin{abstract}
The two-magnon problem for the frustrated XXZ spin-$1/2$
Heisenberg Hamiltonian and external magnetic fields exceeding the
saturation field $B_s$ is considered. We show that the problem can
be \emph{exactly} mapped onto an effective tight-binding impurity
problem. It allows to obtain explicit exact expressions for the
two-magnon Green's functions \emph{for arbitrary dimension and
number of interactions}. We apply this theory to a quasi-one
dimensional helimagnet with ferromagnetic nearest neighbor $J_1<0$
and antiferromagnetic next nearest neighbor $J_2>0$ interactions.
An outstanding feature of the excitation spectrum is the existence
of two-magnon bound states. This leads to deviations of the
saturation field $B_s$ from its classical value
$B_s^{\mathrm{cl}}$ which coincides with the one-magnon
instability. For the refined frustration ratio
$|J_2/J_1|>0.374661$ the minimum of the two-magnon spectrum occurs
at the boundary of the Brillouin zone. Based on the two-magnon
approach, we propose general analytic expressions for the
saturation field $B_s$, confirming known previous results for
one-dimensional isotropic systems, but explore also the role of
interchain and long-ranged intrachain interactions as well as of
the exchange anisotropy.
\end{abstract}

\date{24.10.2006}
\maketitle

\section{Introduction}

We study a spin-$1/2$ quasi-one dimensional helimagnet with ferromagnetic ($%
J_{1}<0$) nearest neighbor and antiferromagnetic ($J_{2}>0,\
|J_{2}/J_{1}|>1/4$) next-nearest neighbor in-chain interactions.
In the classical approximation the spins are vectors. In zero
magnetic field, they form a planar spiral structure (say in the
$xy$ plane) with a pitch angle
$$
\cos \varphi ^{\mathrm{cl}}= -J_1/4J_2
$$
between neighboring spins. When a magnetic field is applied along
the $z$-axis, the spin moments are inclined toward the $z$-axis by
an angle
$$
\sin \theta ^{\mathrm{cl}}=8\mu BJ_2/(4J_2+J_1)^2,
$$
where $\mu =-g\mu _B$ is the value of the magnetic moment. For fields
greater than
\begin{equation}
\label{Bclas}\mu B_s^{\mathrm{cl}}= (4J_2+J_1)^2/8J_2
\end{equation}
the angle $\theta =\pi /2$ and the system becomes ``ferromagnetic'' (fully
polarized uniform state).

In the quantum case, this high-field ferromagnetic state becomes
unstable when the frequency of a certain excitation mode vanishes.
The one-particle instability occurs just at the classical field
$B_s^{\mathrm{cl}}$ given by Eq.\ (\ref {Bclas}). For the
collinear antiferromagnet and the obtuse-angle helimagnet
($\varphi >\pi /2$) the quantum saturation field coincides with
the classical one \cite{Gerhardt98}. In contrast, for an
acute-angle helimagnet ($\varphi <\pi /2$) the corresponding
saturation field \emph{exceeds} the classical value
$B_s^{\mathrm{cl}}$ \cite{Chubukov91, DK05} due to the existence
of $n$ -magnon bound states below the $n$-magnon continuum (see
section II.B of Ref. \onlinecite{DK05}).

Below we derive an explicit exact expression for the two-magnon
Green's function at magnetic fields $B>B_s$. It exhibits isolated
poles below the two-particle continuum which correspond to
two-magnon bound states. According to Refs.
\onlinecite{Chubukov91,DK05}, for $|J_2/J_1|>|\alpha _c|\approx
0.38$ the two-magnon spectrum minimum determines the saturation
field $B_s>B_s^{\mathrm{cl}}$. Our approach allows to refine the
value of $\alpha _c$, to reproduce their results for $B_s$ for the
isotropic $J_1$-$J_2$ Heisenberg model and to generalize it to
more complex situations of exchange anisotropy and interchain
interaction as well as of an additional in-chain interaction
$J_3$.


\section{The model and notations}

The Hamiltonian of the model reads
\begin{equation}
\label{H}\hat{H} = -\mu B\sum_m\hat{S}^{z}_m+\frac{1}{2}\sum_{m,r}%
\left[J^z_{r}\hat{S}^{z}_m\hat{S}^{z}_{m+r}+ \frac{J_{r}^{xy}}{2}\left(\hat{S%
}^{+}_m\hat{S}^{-}_{m+r}+\hat{S}^{-}_m\hat{S}^{+}_{m+r}\right)\right],
\end{equation}
where $m$ enumerates the sites in the chain, $r$ determines the nearest ($%
r=\pm 1$) and the next-nearest ($r=\pm 2$) neighboring sites. We
have allowed for an uniaxial anisotropy of the exchange
interactions. We restrict ourselves to the case of $s=1/2$. Then
the model given by Eq.\ (\ref{H}) can be applied to undoped
edge-shared chain cuprates\cite{Drech06}. Here the spin operators
$\hat S_m^\alpha $ may be expressed via the hard-core boson
operators $b_m$
\begin{eqnarray}  \label{hardc}
\hat{S}^{+} &\equiv & b,\ \hat{S}^{-}\equiv b^{\dagger },\ \hat{S}^{z}\equiv
\frac{1}{2}- \hat{n}, \\
\label{bcomut}\left[ b_{m},b_{m^{\prime }}^{\dagger }\right] &=& \left( 1-2\hat{n}%
_{m}\right) \delta _{mm^{\prime }},\    \\
\label{nconstr}\hat{n} &=& b_{m}^{\dagger }b_{m}=0,1,
\end{eqnarray}
where the square brackets stand for the commutator, and $m$
denotes the site index. The ferromagnetic state corresponds to the
vacuum state $b\left| FM\right\rangle =b\left| 0\right\rangle =0$.
Then the Hamiltonian (\ref{H}) can be rewritten as
\begin{eqnarray}
\hat{H} &=& \hat{H}_{0}+\hat{H}_{int},  \label{Hviab} \\
\hat{H}_{0} &=& \omega _{0}\sum_{m}\hat{n}_{m}+ \frac{1}{2}%
\sum_{m,r}J_{r}^{xy}b_{m}^{\dagger }b_{m+r},  \label{H0} \\
\omega _0 &\equiv & \mu B-\frac 12\sum_rJ_r^z, \nonumber \\
\hat{H}_{int} &=& \frac{1}{2}\sum_{m,r}J^z_{r}\hat{n}_{m}\hat{n}_{m+r}.
\label{Hint}
\end{eqnarray}
The transverse part of $\hat H$ (\ref{H}) defines the one-particle
hoppings in $\hat H_0$ (\ref{H0}), the Ising part contributes the
interaction (\ref{Hint}) and on-site energy value $\omega _0$.

We shall study the one- and two-particle excitation spectra of the
Hamiltonian given by Eq.\ (\ref{Hviab}) which will be obtained
from the singularities (poles and branch cuts) of the
corresponding retarded Green's functions (GF)
\begin{eqnarray}
G^{(1)}(q,\omega ) &=&\left\langle \left\langle b_{q}|b_{q}^{\dagger
}\right\rangle \right\rangle ,  \label{G1} \\
G_{l,a}(k,\omega ) &=&\left\langle \left\langle A_{k,l}|A_{k,a}^{\dagger
}\right\rangle \right\rangle ,  \label{Gla}
\end{eqnarray}where
\begin{eqnarray}
\langle \langle \hat{X}|\hat{Y}\rangle \rangle &\equiv &-\imath
\int_{t^{\prime }}^{\infty }\!\!dte^{i\omega (t-t^{\prime })}\left\langle %
\left[ \hat{X}(t),\hat{Y}(t^{\prime })\right] \right\rangle ,  \nonumber \\
\hat{A}_{k,l} &\equiv &\frac{1}{\sqrt{N}}\sum_{q}\mathrm{e}%
^{iql}b_{k/2+q}b_{k/2-q}  \nonumber \\
&=&\frac{1}{\sqrt{N}}\sum_{m}\mathrm{e}^{-ik(m+l/2)}b_{m}b_{m+l}.
\label{Akl}
\end{eqnarray}

The expectation value $\langle \ldots \rangle$ denotes the ground
state average, the time dependence of an operator $\hat{X}(t)$ is
given by $\hat{X}(t)=\mathrm{e}^{it\hat{H}}
\hat{X}\mathrm{e}^{-it\hat{H}}$, and the Fourier transform of
$b_m$ reads $b_{q}=N^{-1/2}\sum_{m}\exp (-\imath qm)b_{m}$. $N$
denotes the total number of sites.

\section{The one-magnon spectrum}

The equation of motion for the hard-core boson operators (\ref{hardc}) reads%
\begin{equation}
\imath \frac{d}{dt}b_{m}=\left[ b_{m},\hat{H}\right] =\omega
_{0}b_{m}+\sum_{r}\left[ J_{r}^{xy}\left(
\frac{1}{2}-\hat{n}_{m}\right)
b_{m+r}+J_{r}^{z}\hat{n}_{m+r}b_{m}\right] .
\end{equation}
For the ferromagnetic ground state, the terms proportional to
$\hat{n}$ do not contribute to the one-magnon GF (\ref{G1}). This
means that the usually infinite hierarchy of equations of motion
including higher order Green's function is cutted exactly and
closed rigorous expressions for all n-magnon Green's function can
be obtained in principle. In particular, the one-magnon GF becomes
simply
\begin{equation}
\label{g0}G^{(1)}(q,\omega )=\left\langle \left\langle b_{q}|b_{q}^{\dagger
}\right\rangle \right\rangle =\left( \omega -\omega _{q}^{SW}\right) ^{-1},
\end{equation}
where
\begin{equation}
\label{wSW}\omega _{q}^{SW}=\omega _{0}+\frac{1}{2}\sum_{r}J_{r}^{xy}\exp
\left( \imath qr\right)
\end{equation}
is the free spin-wave dispersion.

The dispersion $\omega _{q}^{SW}$ has a minimum at the helical
wave vector $q_{0}=\varphi ^{\mathrm{cl}}/a$, where $a=1$ is the
lattice constant.

For the anisotropic Hamiltonian (\ref{H}) it is convenient to define
\begin{equation}
\label{aldef}\alpha \equiv J_2^{xy}/J_1^{xy},\ \Delta _i\equiv
J_i^z/J_i^{xy}.
\end{equation}
Then
\begin{equation}
\label{phianiso}\cos \varphi ^{\mathrm{cl}}=-1/4\alpha ,
\end{equation}
and for magnetic fields values smaller than
\begin{equation}
\label{Bclaniso}
\begin{array}{c}
\mu B_s^{
\mathrm{cl}}=J_1^z+J_2^z+J_2^{xy}+\frac{\left( J_1^{xy}\right) ^2}{J_2^{xy}}%
= \\ =J_2^{xy}\left[ \frac{\Delta _1-1}\alpha +\Delta _2-1+\frac{\left(
4\alpha +1\right) ^2}{8\alpha ^2}\right]
\end{array}
\end{equation}
$\omega _{q_0}^{SW}$ becomes negative. Evidently, in the isotropic
case $\Delta _1=\Delta _2=1$, Eq.\ (\ref{Bclaniso}) reduces to
Eq.\ (\ref{Bclas}).

\section{The two-magnon Green's function}

The operator $\hat{A}_{k,l}$ (\ref{Akl}) annihilates a pair of
particles, separated by the distance $l$ and moving with total
quasimomentum $k$. A two-particle bound state manifests itself by
an isolated pole of the two-magnon GF (TMGF) (\ref{Gla}). The
hard-core condition (\ref{nconstr})
demands $\hat{A}_{k,0}\equiv 0$. We see also from Eq.\ (\ref{Akl}) that $\hat{A%
}_{k,l}=\hat{A}_{k,-l}$ . The time evolution of $\hat{A}_{k,l},\ l>0$ is
given by the relation%
\begin{eqnarray}
\imath \frac{d}{dt}\hat{A}_{k,l} &=&\left[ \hat{A}_{k,l},\hat{H}\right]
=2\omega _{0}\hat{A}_{k,l}+  \label{dAdt} \\
&&\frac{1}{\sqrt{N}}\sum_{m}\mathrm{e}^{-ik(m+l/2)}\left\{ \sum_{r}\left[
J_{r}^{xy}\left( \frac{1}{2}-\hat{n}_{m}\right) b_{m+r}+J_{r}^{z}\hat{n}%
_{m+r}b_{m}\right] b_{m+l}+\hat{A}_{m,l}^{\prime }\right\} , \\
\hat{A}_{m,l}^{\prime } & \equiv & \sum_{r}b_{m}\left[ J_{r}^{xy}
\left( \frac{1}{2}-\hat{n}_{m+l}\right)
b_{m+l+r}+J_{r}^{z}\hat{n}_{m+l+r}b_{m+l}\right] . \nonumber
\end{eqnarray}Using the commutation relations (\ref{bcomut}), and the
symmetry $J_{-r}=J_{r}$ we rewrite the operator $\hat{A}_{k,l}^{\prime }$ in
the normal form%
\begin{eqnarray*}
\hat{A}_{m,l}^{\prime } &=&\sum_{r}\left[ J_{r}^{xy}\left(
\frac{1}{2}-\hat{n}_{m+l}\right) b_{m}b_{m+l+r}+J_{r}^{z}\hat{n}_{m+l+r}b_{m}b_{m+l}\right] +
\\
& + &\sum_{r}\left[ -\delta _{l,0}J_{r}^{xy}b_{m}b_{m+l+r}+\delta
_{l,r}J_{r}^{z}b_{m}b_{m+l}\right] .
\end{eqnarray*}
Again we note that for the ferromagnetic ground state, the
terms containing the operators $\hat{n}$ do not contribute to the
GF and as discussed in the previous section corresponding higher
order GF vanish exactly. Then, within the subspace of two-particle
excitations above the ferromagnetic ground state, we may write
rigorously
\begin{equation}
\label{AH}\left[ \hat{A}_{k,l},\hat{H}\right] =\left( 2\omega
_{0}+J_{l}^{z}\right) \hat{A}_{k,l}+\left( 1-\delta _{l,0}\right)
\sum_{r}J_{r}^{xy}\cos \frac{kr}{2}\hat{A}_{k,l+r}.
\end{equation}
Thus, the problem of calculation of the TMGF(\ref{Gla}) is equivalent to the
impurity problem for the one-dimensional tight-binding-like Hamiltonian
\begin{eqnarray}
\hat{H}_{tb}(k) &=&\hat{T}+\hat{V},  \label{Htb} \\
\hat{T} &=&2\omega _{0}\sum_{m}\left\vert m\right\rangle \left\langle
m\right\vert +\sum_{m,r}\left\vert m+r\right\rangle t_{r}\left\langle
m\right\vert ,  \nonumber \\
\hat{V} &=&\sum_{m^{\prime }}\left\vert m^{\prime }\right\rangle \varepsilon
_{m^{\prime }}\left\langle m^{\prime }\right\vert ,  \nonumber
\end{eqnarray}where
\begin{equation}
t_{r}(k)=J_{r}^{xy}\cos \frac{kr}{2},\ m^{\prime }=0,r,\
\varepsilon _{0}=\infty ,\ \varepsilon _{r}=J_{r}^{z}.
\end{equation}
Let us note that the infinite value of $\varepsilon _{0}$ is the
result of the hard-core constraint given by Eq.\ (\ref{nconstr}).
The periodic part $\hat{T}$ results from $\hat{H}_{0}$ (\ref{H0}),
and $\hat{H}_{int}$ (\ref{Hint}) defines the changes of the
on-site energies on impurity sites.

It is easy to see that
\begin{equation}
\label{Gfi}G_{l,a}(k,\omega )=\left\langle \phi _l\right| \left( \omega -%
\hat H_{tb}\right) ^{-1}\left| \phi _a\right\rangle ,
\end{equation}
where $\left| \phi _j\right\rangle =\left( \left| j\right\rangle
+\left| -j\right\rangle \right) /\sqrt{2}$, $j=l,a$.

In a standard way, we will use the identity
\begin{eqnarray}
\left( \omega -\hat{H}_{tb}\right) ^{-1} &=&\left( \omega -\hat{T}\right)
^{-1} \\
&+&\left( \omega -\hat{T}\right) ^{-1}\hat{V}\left( \omega -\hat{H}%
_{tb}\right) ^{-1}  \nonumber
\end{eqnarray}for the solution of the impurity problem in the real space.
After some algebra we obtain
\begin{equation}
\label{J1J2res}G_{1,1}(k,\omega )=\frac{1}{\displaystyle-J_{1}^{z}+\frac{1}{%
\displaystyle G_{1,1}^{(0)}+\frac{G_{1,2}^{(0)}J_{2}^{z}G_{2,1}^{(0)}}{%
1-G_{2,2}^{(0)}J_{2}^{z}}}},
\end{equation}
where $G_{l,a}^{(0)}$ is the GF of non-interacting hard-core bosons
($\hat{H}=\hat{H}_{0}$)
\begin{eqnarray}
G_{l,a}^{(0)}(k,\omega ) &=&\left\langle \left\langle
A_{k,l}|A_{k,a}^{\dagger }\right\rangle \right\rangle _{0}=    \nonumber \\
&=&g_{l+a}+g_{l-a}-\frac{2g_{l}g_{a}}{g_{0}},  \label{Gfi0} \\
g_{l}(k,\omega ) &\equiv &  \nonumber \\
&\frac{1}{N}&\sum_{q}\frac{\cos ql}{\omega -\left( \omega
_{k/2+q}^{SW}+\omega _{k/2-q}^{SW}\right) }. \label{gl}
\end{eqnarray}

\section{The two-magnon bound states and the saturation field}

In the derivation of the exact expression for the Green's function
(\ref{J1J2res})-(\ref {gl}) we have used the mathematical
equivalence of the Heisenberg model (\ref{H}) at high fields with
the 1D impurity problem (\ref{Htb}). But the obtained explicit
expressions have a rather complicated form. Fortunately, the
physics of the same 1D impurity problem helps also considerably in
its further analysis.

The branch cut of $G_{1,1}(k,\omega )$ (\ref{J1J2res}) is defined
by the continuous part of the spectrum of $\hat{H}_{tb}(k)$ given
by Eq.\ (\ref{Htb}) that corresponds to the two particle continuum
of the starting Hamiltonian (\ref{H}). Its boundaries may be found
from the spectrum of the periodic part $\hat{T}$
$$
E(k,q)=2\left[ \omega _{0}+t_{1}(k)\cos q+t_{2}(k)\cos 2q\right]
=\omega _{k/2+q}^{SW}+\omega _{k/2-q}^{SW}.
$$
Since $t_{1}(k)<0$ for all $k$, we have
\begin{eqnarray}
E(k,q_{1}) &<&E(k,q)<E(k,\pi ),\ |k|<k_{1},  \label{contb} \\
E(k,0) &<&E(k,q)<E(k,\pi ),\ k_{1}<|k|<k_{2}, \\
E(k,0) &<&E(k,q)<E(k,q_{1}),\ k_{2}<|k|<\pi
\end{eqnarray}where
\begin{eqnarray*}
\cos q_1 &=& -t_1(k)/4t_2(k) \\
E(k,q_{1}) &\equiv &2\omega _{0}-t_{1}^{2}/4t_{2}-2t_{2}, \\
k_{1} &\equiv &2\arccos \frac{\sqrt{128\alpha ^{2}+1}+1}{16|\alpha
|}<\pi /2, \quad t_1(k_1)/4t_2(k_1)=-1,
\\
k_{2} &\equiv &2\arccos \frac{\sqrt{128\alpha ^{2}+1}-1}{16|\alpha
|}>\pi /2, \quad t_1(k_2)/4t_2(k_2)=1.
\end{eqnarray*}

As we will see below, the point $k=\pi $ has a special meaning.
For this value of $k$ the nearest-neighbor hopping $t_{1}(\pi
)=0$, and the Hamiltonian $\hat{T}$ describes two non-interacting
linear chains (i.e.\ the sites with odd or even numbers $m$) with
a hopping $t_{2}(\pi )=-J_{2}^{xy}$ inside each chain. The account
of the hard-core constraint (\ref{nconstr}), i.e. the
introduction of $\hat{V}_{0}=|0\rangle \varepsilon _{0}\langle 0|$ with $%
\varepsilon _{0}=\infty $ entering the Hamiltonian
$\hat{H}_{tb}(k)$ given by Eq.\ (\ref{Htb}) , does not influence
the chain with odd sites, but the chain with even sites $m$ is
broken into two independent semi-infinite chains. Now, it is easy
to account for the rest of terms in the impurity Hamiltonian
$\hat{V}$ , because $\varepsilon _{1}=J_{1}$ and $\varepsilon
_{2}=J_{2}$ affect different chains. The GF $G_{2,2}^{(2)}(\pi
,\omega )$ has a particular simple form. It can be obtained e.g.\
by the recursion method
\begin{equation}
\label{G22}G_{2,2}(\pi ,\omega )=\left[ \omega -(2\omega
_{0}+J_{2}^{z})-\left( J_{2}^{xy}\right) ^{2}G_{2,2}^{(0)}(\pi ,\omega
)\right] ^{-1}=\frac{1}{J_{2}^{xy}\left[ z-\Delta _{2}-\tau (z)\right] },
\end{equation}
where  the dimensionless energy
\begin{equation}\label{z}
z\equiv (\omega -2\omega _{0})/J_{2}^{xy}
\end{equation}
 is introduced, and $\tau (z)\equiv \left( z-\sqrt{z^{2}-4}\right)
/2=1/\left[ 1-\tau (z)\right] $ is the local Green's function on
the first site of the unperturbed semi-infinite chain in
dimensionless units. A simple analysis of the expression
(\ref{G22}) shows that besides the branch cut in the interval
$-2<z<2$ of the real axis, $G_{2,2}(\pi ,\omega )$ may have an
isolated pole. The pole exists for $\Delta _{2}>1$ \emph{above}
the continuum, i.e. a bound state exists for the easy-axis
anisotropy of the next-nearest neighbor exchange. The expression
for $G_{1,1}(\pi ,\omega )$ is more complicated than $G_{2,2}(\pi
,\omega )$
\begin{eqnarray}
G_{1,1}(\pi ,\omega ) &=&\left[ \omega -(2\omega
_{0}+J_{1}^{z}-J_{2}^{xy})-J_{2}^{xy}\tau (z)\right] ^{-1}  \label{G11n2} \\
&=&\frac{1}{J_{2}^{xy}\left[ z-\Delta _{1}/\alpha +1-\tau (z)\right] }.
\end{eqnarray}

Note that for any acute angle helimagnet ($\alpha <0$) a bound
state should exist \emph{below} the continuum. Indeed, the
condition $ G_{1,1}^{-1}(k,\omega _b)=0$ gives
\begin{equation}
\label{zb}z_b(\pi )=\frac{\Delta _1}\alpha -1+\frac 1{\frac{\Delta _1}\alpha
-1}<-2.
\end{equation}

For $k\neq \pi $, a bound state exists, too \cite{Chubukov91}. In
Ref.\onlinecite {Chubukov91}, the isotropic version ($\Delta
_1=\Delta _2=1$) of the Hamiltonian (\ref{H}) was considered.
There A.~Chubukov has found that the dispersion of the two-magnon
bound state
 $z_b(k)$ exhibits a minimum at $k=\pi $ for $|\alpha
|>|\alpha _c|\approx 0.38$. The latter number will be refined
below.

Based on extensive numerical work\cite{Cabra00,HMeisner06} for
finite chains, and qualitative discussion in Ref.\
\onlinecite{DK05}, we strongly believe that the two-particle bound
state defined by (\ref{zb}) is the excitation  with the
\emph{lowest} energy per flipped spin in the system for this
parameter regime. The absolute dominance of two-magnon states
manifests itself by $\Delta S^z=2$ steps of the calculated
magnetization curves $M(H)$ at high fields. Only for $\alpha
\lesssim 0.4$ steps with $\Delta S^z=3$ are observed (see Fig.\ 1
of Ref.\ \onlinecite{HMeisner06}). But we admit that from a formal
point of view, the full rigorous solution should also include the
analysis of the problem for the arbitrarily $n$-magnon
 bound-states ($n\geq 3$) in a similar way as done here. However,
the corresponding calculations are rather cumbersome and
particular examples ($n=3,4$) are left for future consideration.

Then the quantum saturation field is determined by the condition
that the two-magnon energy vanishes (i.e. the two-magnon
instability of the field-induced ferromagnetic ground state). This
way, the central result of the present work yields
\begin{eqnarray}
\omega _b(\pi ,B_s) &=& 2\mu (B-B_s)=J_2^{xy}z_b(\pi
)+2\omega _0=0, \label{wb2Bs}\\
&\mbox{or} & \nonumber\\
\mu  B_s &=& \frac 12\sum _r J_r^z-J_2^{xy}z_b(\pi ). \hspace{5
cm} (36^{\prime}) \nonumber
\end{eqnarray}
Which gives explicitly
\begin{eqnarray}
B_s &=& \frac{2J^{xy}_2 (J^z_2+J^{xy}_2)-(J^z_1)^2-2J^z_2 J^z_1}
{2\mu
(J^{xy}_2-J^z_1)}  \label{chubukon} \\
&=& \frac{J^{xy}_2}{2\mu }\left[\frac{2(\Delta _2+1)-(\Delta _1/\alpha
)^2-2\Delta _1/\alpha } {1-\Delta _1/\alpha }\right].  \nonumber
\end{eqnarray}
For the isotropic case, this result was first obtained in Ref.
\onlinecite{Chubukov91} by solving an integral equation which
results from a summation of a  sequence of ladder diagrams. From
our straightforward derivation it is clear that the result is
exact within the adopted two-magnon approach, as also pointed out
in Ref. \onlinecite{DK05}.

In order to find the parameter range where the value of $z_b(\pi
)$ (see Eq.\ (\ref{zb})) yields the minimum of the bound state
dispersion $z_b(k)$ , we consider the expressions (\ref{J1J2res}
)-(\ref{gl}) in the vicinity of $k=\pi $, $z=z_b(k)$. After
straightforward calculations given in the Appendix we obtain
\begin{equation}
\label{zbk}z_b(k)\approx z_b(\pi )+\frac{\left( k-\pi \right) ^2}{2m_{eff}},
\end{equation}
where
\begin{equation}
\label{meff}\frac 1{2m_{eff}(\alpha ,\Delta _1,\Delta _2)}=\frac 12+\frac{%
\Delta _1-\alpha }{4\alpha \Delta _1^2}+\alpha \frac{\alpha -2\Delta _1}{%
2\left( \alpha -\Delta _1\right) ^2}-\frac{(2\alpha -\Delta _1)\Delta _2}{%
4\Delta _1\left( \alpha -\Delta _1\right) \left[ \alpha \left( 1+\Delta
_2\right) -\Delta _1\right] }.
\end{equation}

\begin{figure}[tbp]
\epsfig{file=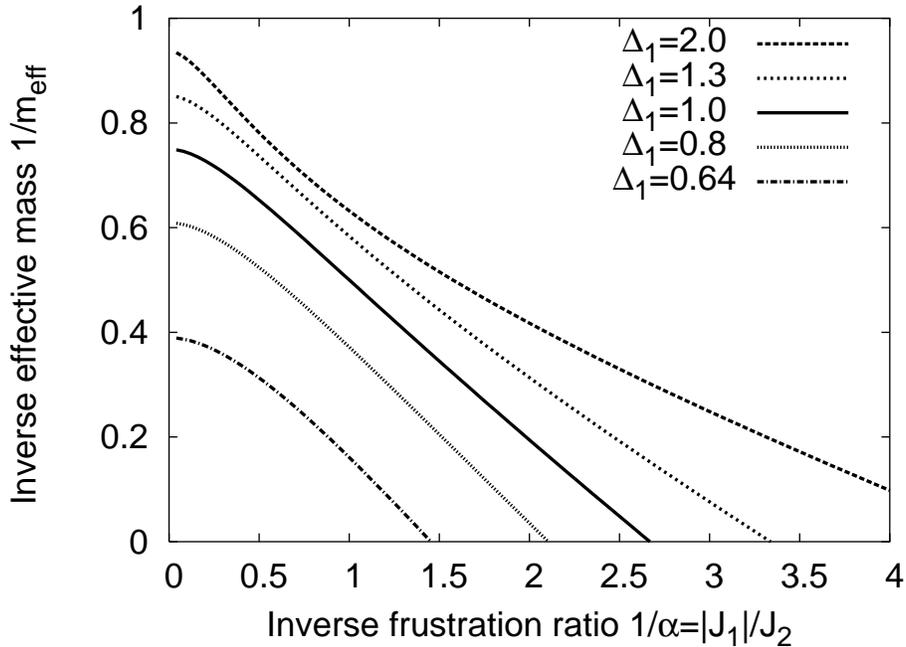} \caption{The inverse effective mass
(\ref{meff}) as a function of frustration ratio and anisotropy
$\Delta _1=J_1^z/J_1^{xy}$ for the
 $J_1-J_2$ model.} \label{fig1}
\end{figure}

\begin{figure}[tbp]
\epsfig{file=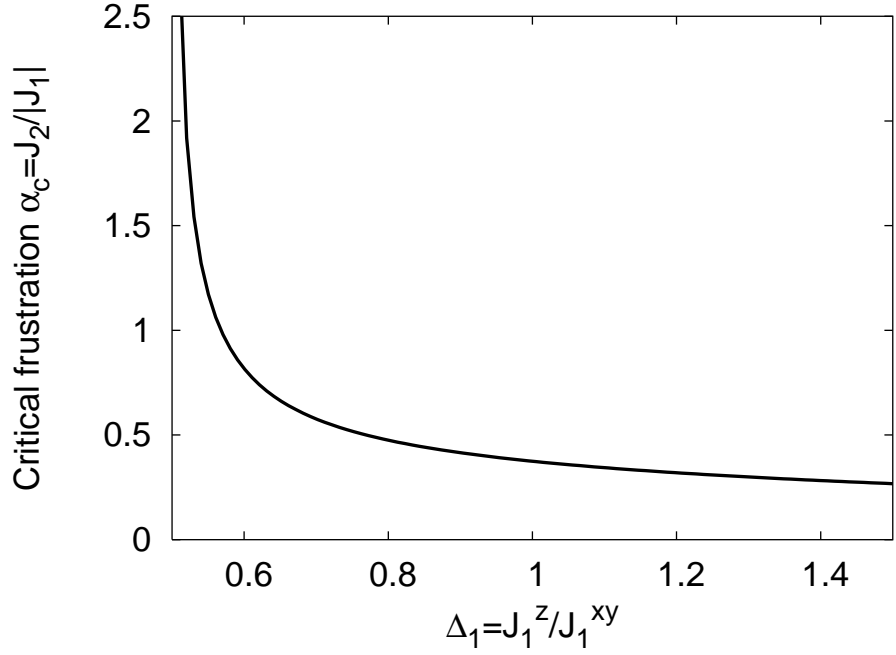} \caption{The dependence of the frustration
ratio value $\alpha _c$, for which the effective mass (\ref{meff})
changes the sign, on anisotropy $\Delta _1=J_1^z/J_1^{xy}$ for the
 $J_1-J_2$ model.} \label{fig2}
\end{figure}

The dependence of the right-hand side of Eq.\ (\ref{meff}) on the
inverse frustration ratio $|J_1/J_2|$ is shown in Fig.\ \ref{fig1}
for different values of the nearest-neighbor exchange anisotropy
$\Delta _1$ . The dependence on the second neighbor exchange
anisotropy is weak in the vicinity of the isotropic point $\Delta
_2=1$. We see that $m_{eff}$ is positive for large values of
frustration $J_2\gg |J_1|$ and changes the sign at some critical
value $ \alpha _c$, where the dispersion minimum is transformed to
a local maximum. For $\Delta _2=1$ the condition
$1/2m_{eff}(\alpha _c,\Delta _1,1)=0$ reduces to a cubic equation
for $\alpha _c$ and we have
\begin{equation}
\label{alfc}\alpha _c(\Delta _1)=\Delta _1\left\{ \frac 23-2R(\Delta _1)\cos
\left[ \frac{\phi (\Delta _1)+2\pi }3\right] \right\} ,
\end{equation}
where
$$
R(\Delta _1)\equiv -\frac 13\sqrt{2\frac{5\Delta _1^2+1}{4\Delta
_1^2-1}},\ \phi (\Delta _1)\equiv \arccos \left( -\frac
7{54R^3}\right).
$$
The critical frustration dependence on anisotropy is shown on
Figure \ref{fig2} . In the isotropic case, we have $R(1)=-2/3$,
$\phi (1)=\arccos \left( 7/16\right) $, and
\begin{equation}\label{alfc1}
\alpha _c(1)=\frac{2}{3}\left(2\cos\frac{\phi
(1)+2\pi}{3}+1\right)\approx -0.37466105983527,
\end{equation}
which refines $|\alpha _c(1)|\approx 0.38$ calculated before
\cite{Chubukov91,DK05}. If $\Delta _1\rightarrow 0.5$, then
$\alpha _c$ from Eq.\ (\ref{alfc}) diverges, i.e.\ the effective
mass $m_{eff}(\Delta _1=0.5)$ becomes negative for all frustration
values.

\begin{figure}[tbp]
\epsfig{file=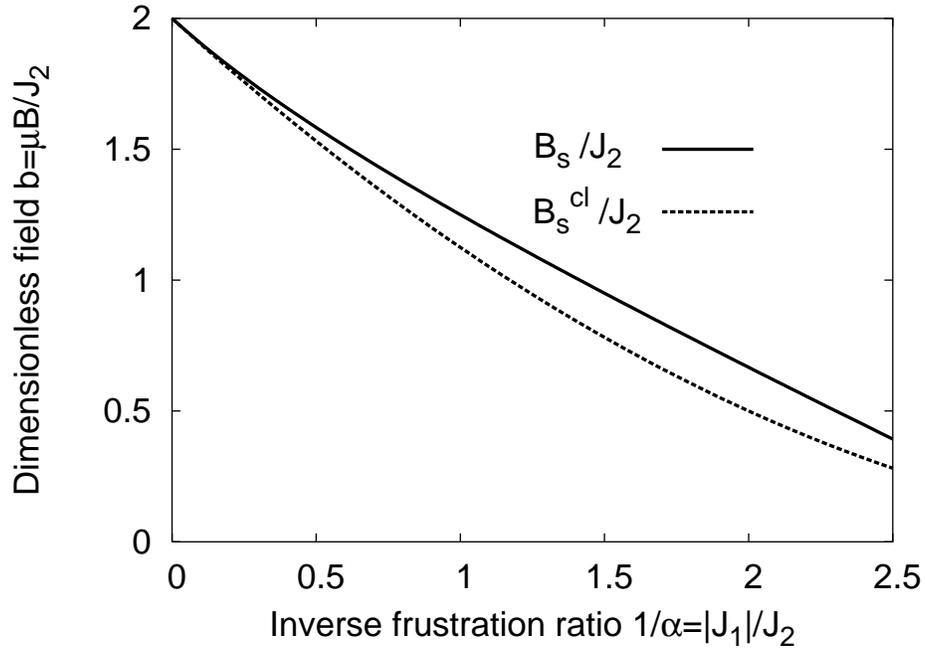} \caption{The two-particle ($\protect\mu
B_{s,2}/J_2$ - solid line) and one-particle ($\protect\mu
B_{s,2}^{\mathrm{cl}}/J_2$ - dashed line) values of the saturation
field as a function of frustration ratio $|J_1|/J_2$ for the
isotropic $J_1-J_2$ model.} \label{fig3}
\end{figure}

The Figure \ref{fig3} shows the quantum and classical saturation
field dependencies on the parameter values of the 1D isotropic
$J_1-J_2$ model. We have chosen $J_2$ as the unit of energy. We see
that the quantum effect is most pronounced for frustration values
$|J_2/J_1|\sim 1$.

In the region $0.25<|\alpha |<|\alpha _c|$ the minimum $z_b(k)$
for the isotropic model shifts into the point\cite{Chubukov91}
$k=2q_0=2\arccos (-1/4\alpha )$. The authors of Ref.
\onlinecite{DK05} argue that in this case the saturation field is
determined by bound states of three and/or more
magnons\cite{DK05}; such a situation is out of the scope of this
paper. For the anisotropic $J_1-J_2$ model, there is a third
possible scenario. The one- and two-particle instabilities occurs
at different $k$-points. Then, it is possible to have the minimum
of $z_b(k)$ at $k=\pi $ , which is higher in energy than the
lowest boundary of the continuum (\ref{contb})
$z_c(2q_0)=z_c(2q_0)$. This happens e.g.\ for the easy-plane
nearest-neighbor anisotropy values
$$
\Delta _1<\Delta _{1,a}=\frac{1+\sqrt{1+16\alpha ^2}}{8|\alpha |},\ \Delta
_2=1
$$

\begin{figure}[tbp]
\epsfig{file=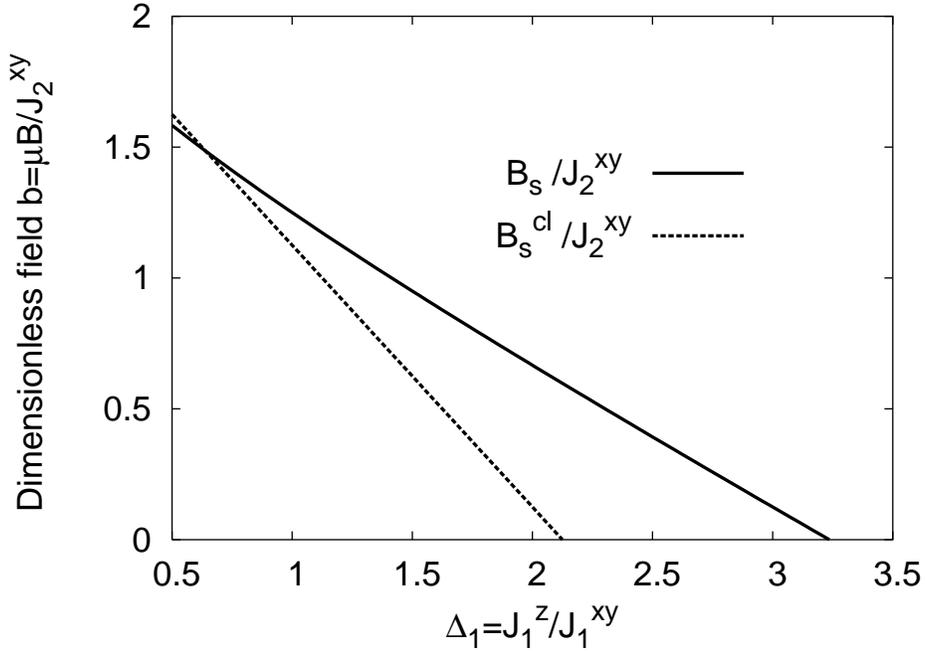} \caption{The quantum ($\protect\mu
B_{s,2}/J_2$ - solid line) and classical ($\protect\mu
B_{s,2}^{\mathrm{cl}}/J_2$ - dashed line) saturation fields for the
anisotropic $J_1-J_2$ model, $J_1=J_2=\Delta _2=1$.} \label{fig4}
\end{figure}

In Fig.\ \ref{fig4} the dependence of the saturation fields on the
anisotropy of the $J_1$ exchange is shown. The former is important
for edge-shared cuprates \cite{aniso}. We see that for $\Delta
_1=(1+\sqrt{17})/8\approx 0.64039$ the lines $B_s(\Delta _1)$ and
$B_s^{\mathrm{cl}}(\Delta _1)$ do intersect. At the same time
$1/2m_{eff}(1,0.64039,1)\approx 0.16178>0$. The unexpected at
first glance result that the classical curve apparently reaches,
then overwhelms the quantum result can be explained by the reduced
attractive ferromagnetic interaction due to the anisotropy, i.e.\
a weakening of the two-magnon "glue". Below this value of $\Delta
_1$ the saturation field is determined by the one-particle
instability, like in the XY model ($\Delta _1=\Delta _2=0$). Thus,
the intersection is not related to a strange quantum versus
classical behaviour, but to the competition between one- and
two-particle instabilities.

It is interesting that a strong easy-axis anisotropy can diminish
the saturation field, and at the point
$$
\Delta _{1,0}=(-\alpha )\left[ 1+\sqrt{3+2\Delta _2}\right]
$$
the field $B_s$ vanishes, i.e. the system's ground state becomes
ferromagnetic \cite{Chubukov91}. Note also the region
$$
\Delta _{1,0}^{\mathrm{cl}}=(-\alpha )\left[ 1+\Delta
_2+\frac{1}{8\alpha ^{2}}\right] <\Delta _1<\Delta _{1,0}
$$
where $B_s^{\mathrm{cl}}=0,\ B_s\neq 0$. Here the classical fully
polarized state is destroyed by quantum fluctuations. A possible
ground state for the system in this parameter regime may be a
collinear state with period 4 described in Ref. \onlinecite{DK05},
or a dimer nematic state, predicted in Ref.
\onlinecite{Chubukov91} for the isotropic model in a high field.

\vspace{0.5 cm}

\section{Additional interactions}

An advantage of our approach is the possibility to apply it to
more complex situations which occur naturally when real chain
compounds are considered.

Indeed, it is easy to realize that the exact mapping of the
two-magnon problem onto the effective tight-binding Hamiltonian
(\ref{Htb}) is not restricted to 1D and to the $J_1-J_2$ model. We
may generalize the Hamiltonian $\hat{H}$ (\ref{Hviab}) including
into summation over $r$ more distant neighbors in chain direction.
This will introduce additional impurities and hoppings in the
effective Hamiltonian (\ref{Htb}). Moreover, we may consider also
2D or 3D systems. Then, the site indices $m$ as well as $r$ in
(\ref{H0}), (\ref{Hint}) becomes vectors with corresponding
changes in the effective Hamiltonian (\ref{Htb}). It is
straightforward to obtain the TMGF (\ref{Gfi}), but the expression
becomes cumbersome. Here we will apply our general approach to
answer the question how is the quantum effect for the saturation
field in the $J_1-J_2$ model modified by some additional
interactions often present in real compounds.

First, we include a small third neighbor in-chain interaction
$J_3$. Such a term may appear as a result of the spin-phonon
interaction in the antiadiabatic regime, when the exchange
constants $J_i\ll \hbar\omega _{ph}$ the characteristic phonon
frequencies, and the spin-phonon interaction is
strong\cite{Fehske99}. It is expected to be small $J_3\ll |J_1|,
J_2$ and antiferromagnetic\cite{Fehske99,Horsch05}. Below, the
subscript 2(3) refers to the $J_1-J_2$ and the $J_1-J_2-J_3$ model
respectively. In this section, for the sake of simplicity we give
only formulae for the isotropic case $J=J^z=J^{xy}$. The minimum
of the one-magnon spectrum (\ref{wSW}) gives the value of the
helicoidal wave vector and the classical value of the saturation
field
\begin{eqnarray}
\cos q_{0,3} &=&\frac{-J_2+\sqrt{J_2^2-3J_3(J_1-3J_3)}}{6J_3} \\
&\approx&\cos q_{0,2}\left(1-\frac{3J_3}{J_1}\right), \\
B_{s,3}^{\mathrm{cl}}&\approx&B_{s,2}^{\mathrm{cl}}+ \frac{J_3}{\mu }\left(1-%
\frac{3}{4\alpha }+\frac{1}{16\alpha ^3}\right)
\end{eqnarray}
we recall that $\alpha <0$

The two-magnon Green's function (\ref{G11n2}) has the form
\begin{equation}
\label{G11n3}G_{1,1}(\pi ,\omega )= \left[\omega -(2\omega _0+J_1-J_2)-
\frac{J_2^2}{\omega -(2\omega _0+J_3)+J_2\tau (\frac{2\omega _0-\omega }{J2})%
}\right]^{-1}.
\end{equation}
For small $J_3$ values, the bound state energy and the saturation
field varies linearly with $J_3$
\begin{eqnarray}
\omega _{b,3}\approx \omega _{b,2}-J_3\left[\frac{\alpha (2-\alpha
)}{(1-\alpha )^4}+1\right], \nonumber\\
B_{s,3}\approx B_{s,2}+\frac{J_3}{2\mu } \left[1+\frac{\alpha
(2-\alpha )}{(1-\alpha )^4}\right], \label{z3}
\end{eqnarray}
where the values $\omega _{b,2}$, and $B_{s,2}$ are given by Eq.
(\ref {chubukon}). The slope of $B_{s,3}^{\mathrm{cl}}$ dependence
on $J_3$ is larger than for $B_{s,3}$. It means, that positive
$J_3$ suppress the quantum effect. The difference of saturation
fields in quantum and classical cases becomes smaller.

As the simplest example for a two-dimensional system we consider a
2D set of chains parallel to the $x$-axis coupled in perpendicular
direction with the strength $J_{\perp }$. Then the one-magnon
dispersion becomes two-dimensional
\begin{equation}
\label{w2D}\omega ^{SW}_{\mathbf{q},2D}=\omega
^{SW}_{q_x,1D}+J_{\perp }\left(\cos q_y a-1\right).
\end{equation}

From this expression one readily obtains
\begin{equation}\label{Bscl2D}
\mu B^{\mathrm{cl}}_{2D}=\mu B^{\mathrm{cl}}_{1D}+J_{\perp }+\mid
J_{\perp }\mid ,
\end{equation}
i.e.\ in this approximation the ferromagnetic interchain
interaction does not affect the saturation field, whereas in the
antiferromagnetic case it is enhanced by $2J_{\perp }$. In
general, such a correction is especially important near the
quantum critical point for ferromagnet-helimagnet transition
$\alpha \approx \left(4+9J_3/J_2\right)^{-1}$, where the 1D
saturation field by definition vanishes. Eq.\ (\ref{Bscl2D})
should be understood as a lower bound for the saturation field
near the critical point. The account of quantum fluctuations will
lead to slightly higher values of $B_s$ according to Ref.\
\onlinecite{DK05}.

\begin{figure}[tbp]
\epsfig{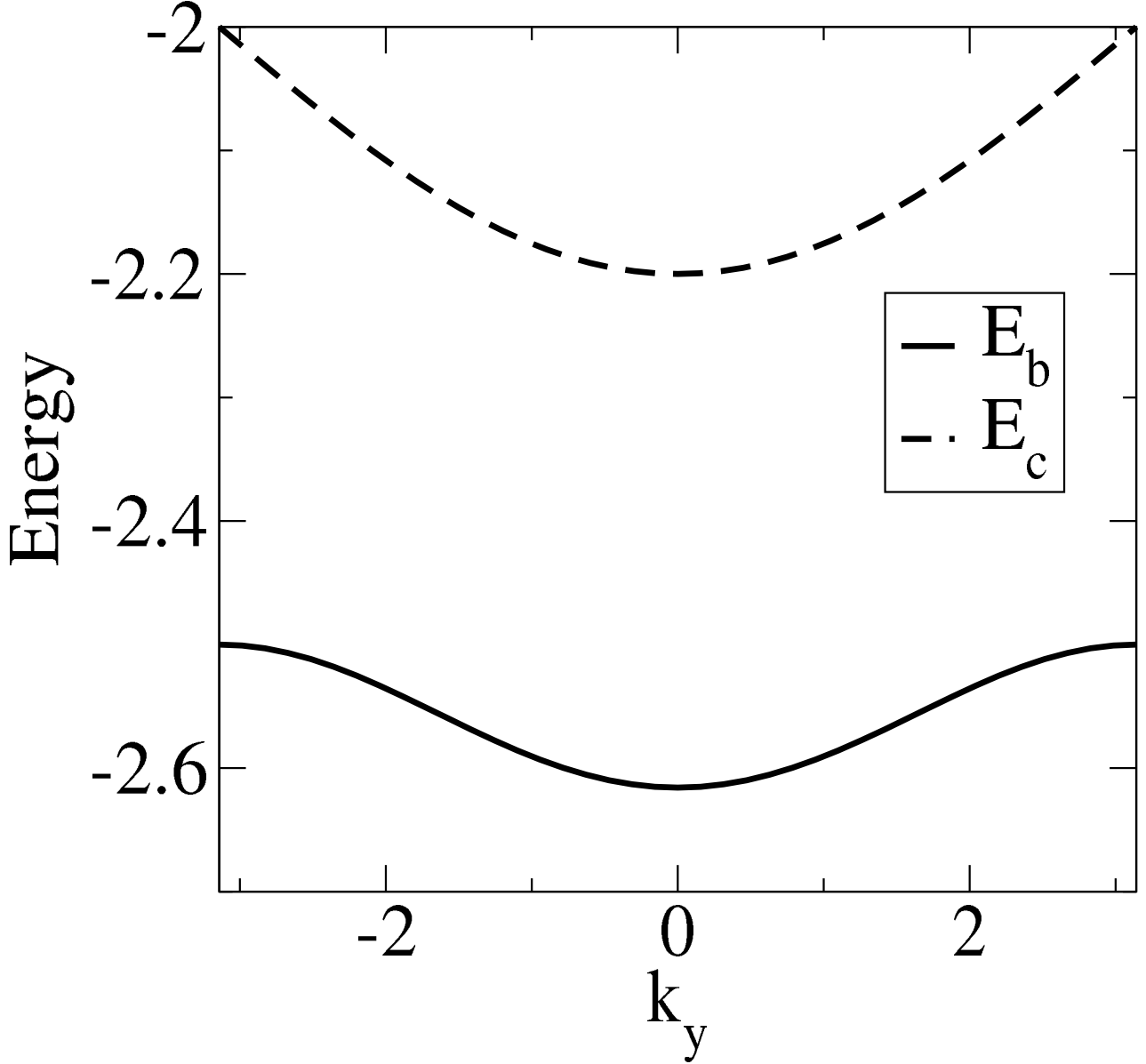} \caption{The two-magnon bound state energy
$z_b(\protect\pi ,k_ya)=(\protect\omega _b(\protect\pi
,k_ya)-2\protect\omega _0)/J_2$ (solid line) and the boundary of
the two-magnon continuum $z_c(\protect\pi ,k_ya)$ (dashed line) as
a function of the quasi-momentum value in the $y$ direction for
$|J_1|=J_2 $, the interchain interaction is chosen as $J_{\perp
}=0.1J_2$, $J_2$ being the unit of energy. For these parameters,
the absolute minimum of continuum is $z_c(2q_0 ,k_ya)=-2.45
J_2>z_b$ } \label{fig5}
\end{figure}

\begin{figure}[tbp]
\epsfig{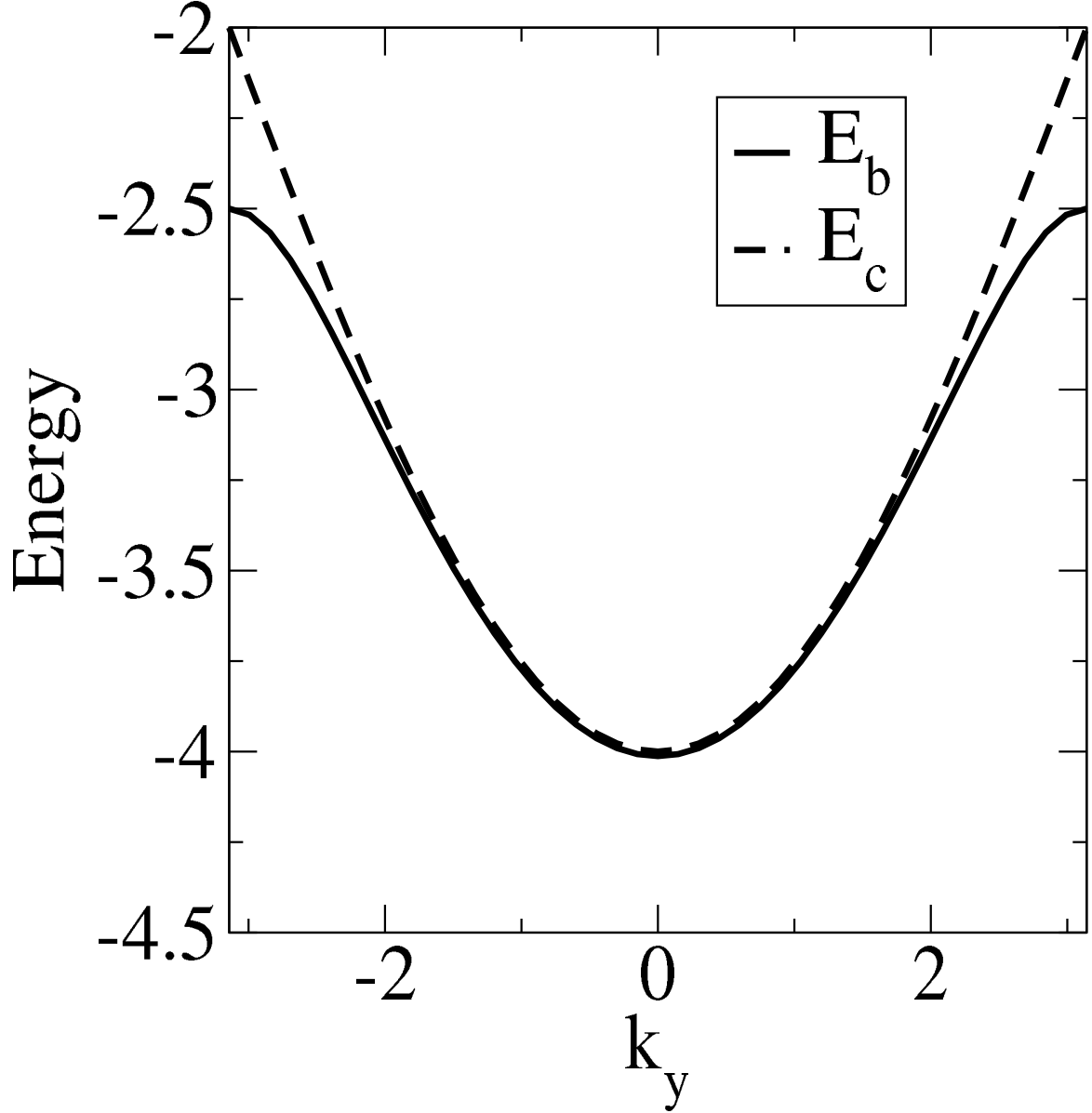} \caption{The same as in previous figure for
$|J_1|=J_2=J_{\perp }$. The absolute minimum of continuum is
$z_c(2q_0 ,k_ya)=-4.25 J_2 < z_b$ } \label{fig6}
\end{figure}

For an arbitrary $\mathbf{k}$-point, the GF (\ref{Gfi}) is found
from the solution of a system of three linear equations, but along
the line $\mathbf{k}=(\pi /a,k_y)$ the system reduces to a single
equation which gives
\begin{equation}
\label{G2D}G_{1,1}(\mathbf{k},\omega )=\left\{\left[G_{1,1}^{(0)}(\mathbf{k}%
,\omega )\right]^{-1}-J_1\right\}^{-1},
\end{equation}
where the two-dimensional spectrum (\ref{w2D}) should be used in
the expression for the non-interacting GF (\ref{Gfi0}). The
dispersion of the isolated pole and the two-particle continuum
boundary are shown in Figs.\ \ref{fig5} and \ref{fig6}. For small
$J_{\perp }$ (Figure \ref{fig5}) one observes a well separated
bound state. Here, the absolute minimum of the continuum occurs at
$\mathbf{k}=(2q_0,0)$ and at $\mathbf{k}=(0,0)$ its energy
$z_c=-2.45 J_2$ exceeds the minimum of bound state dispersion
given by $z_b(0,\pi )=-2.61576 J_2$. For strongly coupled chains
such with $J_{\perp }=|J_1|=J_2$, the pole position becomes very
shallow (Figure \ref{fig6}) and it becomes clear that such a local
minimum exceeds the minimum given by two independent (one-magnon)
excitations ($z_c=-4.25 J_2$ for the parameter set shown in the
caption of Fig.\ \ref{fig6}). In the general 3D problem, one may
expect  that even the bound state itself may disappear.

At variance with the classical case given by Eq.\ (\ref{Bscl2D}),
the solution of Eq.\ (\ref{G2D}) yields for $\mid J_{\perp}\mid
\ll J_2$
\begin{equation}\label{Bs2D}
\mu B_s = \mu B_{s,1D}+J_{\perp}+O(J_{\perp}^2/J_2),
\end{equation}
i.e.\ in this case the saturation field is sensitive to both sign
of interchain interaction. With the increase of $J_{\perp}$ one
finally reaches a critical value, where $z_b=z_c$ and beyond the
"one-magnon" derived Eq.\ (\ref{Bscl2D})  should be used instead
of a "two-magnon" one like Eq.\ (\ref{Bs2D}). Thus, the quantum
effects are maximally pronounced in the 1D case, just as the
localization for the equivalent impurity problem.

More complex interchain interactions derived from band structure
calculations and inelastic neutron scattering data\cite{Enderle05}
and an application to chain cuprates will be given elsewhere.

\section{Conclusion}

We have shown that the internal motion of a two-magnon pair on a
ferromagnetic background is equivalent to the motion of a single
particle described by an effective tight-binding Hamiltonian. This
Hamiltonian is not translationally invariant. It models the
hard-core boson constraint (\ref{nconstr}) by an infinite on-site
energy at the site with zero coordinates,
and each exchange $J_{\mathbf{r}}$ introduces the on-site energy $%
\varepsilon _{\mathbf{r}}$ and the hopping term $t_{\mathbf{r}}=J_{\mathbf{r}%
}\cos \mathbf{kr}/2$. Remarkably, this mapping procedure can be applied to
problems at arbitrary dimension.

The two-magnon Green's function is found exactly by analogy with
the impurity problem. The two-magnon excitation spectrum is found
from poles and branch cuts of the Green's function. For the
quasi-one-dimensional helimagnet with ferromagnetic nearest
neighbor and antiferromagnetic next-nearest neighbor interactions
a bound state of magnons exists. This leads to deviation of the
quantum saturation field $B_s$ from the classical value.

The derived expression for the saturation field $B_s$ (exact
within the two-magnon approach) provides a constraint for
competing exchange interactions. Such a constraint may be useful
in fitting thermodynamic properties such as the magnetic
susceptibility $\chi (T)$ and the magnetic specific heat $c_p(T)$.
In general, high-field magnetization measurements $M(H,T)$ yield
an important information concerning the exchange integrals in
novel materials. Combined with the analysis of other experimental
data, this knowledge may be very helpful to elucidate the relevant
microscopic model for an acute-angle helimagnetic system (i.e.\
having a pitch angle $\leq \pi /2$ at zero magnetic field).

In this work we studied the lowest energy of excited states, i.e.\
the position of the isolated TMGF poles. The obtained Green's
function given by Eq.\ 
(\ref{Gla}) contains the information about the whole
spectrum that is necessary for the calculations of physical
properties for concrete materials. Various application to
edge-shared compounds will be considered elsewhere.

\section*{Acknowledgements}

The authors thank the Heisenberg-Landau program and the DFG
(project 436/UKR/17/8/05) for support. R.O.K. thanks the IFW
Dresden, where the main part of this work has been carried out,
for the hospitality. Discussions with V.Ya. Krivnov, D.V.
Dmitriev, A.V.Chubukov, J. Richter, N.M. Plakida and H. Eschrig
are gratefully acknowledged.

\appendix

\section{}

Here we give details in derivating the Eqs.\
(\ref{zbk}) and (\ref{meff}).

The energy of an isolated pole of GF given by Eq.\ 
(\ref{J1J2res}) is the root of the
equation
\begin{equation}
\label{epole}-J_1^z+\frac 1{\displaystyle G_{1,1}^{(0)}+\frac{%
G_{1,2}^{(0)}J_2^zG_{2,1}^{(0)}}{1-G_{2,2}^{(0)}J_2^z}}=0.
\end{equation}
It depends on the $k$ value via the dependence of the hopping parameters
$t_r$ in $ \hat T$ (\ref{Htb}). We denote $\kappa \equiv \pi -k$,
and expand the left-hand-side of Eq.\ (\ref{epole}) up to terms
$\propto \kappa ^2$. Then
$$
t_1=J_1^{xy}\sin \frac \kappa 2\approx J_1^{xy}\frac \kappa 2,\
t_2=-J_2^{xy}\cos \kappa \approx -J_2^{xy}\left( 1-\frac{\kappa ^2}2\right).
$$

Note that the GF given by Eq.\ (\ref{gl})
$$
g_l(k,\omega )=g_{-l}(k,\omega )=\left\langle R\right| \left( \omega -\hat T%
\right) ^{-1}\left| R+l\right\rangle
 ,
$$
obeys the equation of motion
\begin{equation}
\label{geqm}\left( \omega -2\omega _0\right) g_l=\delta _{l,0}+t_1\left(
g_{l-1}+g_{l+1}\right) +t_2\left( g_{l-2}+g_{l+2}\right) .
\end{equation}
We will calculate $g_0(k,\omega ),g_1(k,\omega )$ directly from Eq.\ 
(\ref{gl}) and use Eq.\ (\ref{geqm}) for the calculation of the other $g_l$
involved into Eq.\ (\ref {epole}). We begin with
\begin{equation}
\label{i0}g_0(k,\omega )=\frac 1{2\pi }\int_{-\pi }^\pi \frac{dQ}{\omega
-2\left( \omega _0+t_1\cos Q+t_2\cos 2Q\right) }.
\end{equation}
The denominator of the integrand is nonzero for $\omega $ outside the
spectrum of $\hat T$. After the substitution $\tau =\tan (Q/2)$ a
straightforward calculations give%
$$
g_0=-\frac 1{8J_2^{xy}\cos \kappa \sqrt{\left( 4q+p^2\right) \left(
q+p-1\right) }}\left[ \frac{p-2+\sqrt{4q+p^2}}{\sqrt{q+1-\sqrt{4q+p^2}}}%
-\right.
$$
$$
\left. -\frac{p-2-\sqrt{4q+p^2}}{\sqrt{q+1+\sqrt{4q+p^2}}}\right] ,
$$
where
$$
p\equiv -\frac{2\sin (\kappa /2)}{4\alpha \cos \kappa },\ q\equiv -\frac{%
z-2\cos \kappa }{4\cos \kappa },\ z\equiv \frac{\omega -2\omega _0}{J_2^{xy}}%
.
$$
Expanding this expression around the point $\left( k=\pi ,\ z=z_b(\pi
)\right) $, we obtain
\begin{equation}
\label{g0ap}g_0\approx -\frac 1{J_2^{xy}\sqrt{z^2-4}}\left[ 1-\kappa ^2(1-%
\frac{\Delta _1}\alpha )^2\frac{4\Delta _1^2\alpha ^2-3\alpha ^2+3\Delta
_1\alpha -\Delta _1^2}{2\Delta _1^3(2\alpha -\Delta _1)}\right] .
\end{equation}
Analogously we obtain%
\begin{eqnarray}
g_1(k,\omega ) &=& \frac 1{2\pi }\int_{-\pi }^\pi \frac{\cos QdQ}
{\omega -2\left( \omega _0+t_1\cos Q+t_2\cos 2Q\right) } \nonumber \\
 &=& -\frac 1{2J_2^{xy}\sqrt{\left( 4q+p^2\right) \left( q+p-1\right)
}}\left[ \frac{2q+p-\sqrt{4q+p^2}}{\sqrt{q+1-\sqrt{4q+p^2}}}-
\frac{2q+p+\sqrt{4q+p^2}}{\sqrt{q+1+\sqrt{4q+p^2}}}\right] \nonumber \\
\label{g1} &\approx & \frac{\kappa \alpha }{2J_2^{xy}}\frac{(\alpha
-\Delta _1)^2 }{\Delta _1^3(2\alpha -\Delta _1)}.
\end{eqnarray}
Now, using Eq.\ (\ref{geqm}), we have%
\begin{equation}\label{g2g0}
g_2+g_0\approx \frac {1-\sqrt{\frac{z-2}{z+2}}}{2J_2^{xy}}\left[
1-\kappa ^2\frac{2\alpha \Delta _1^2\left( 2\alpha
^2-4\Delta _1\alpha +\Delta _1^2\right) -3\left( \alpha -\Delta _1\right) ^3%
}{4\Delta _1^4(2\alpha -\Delta _1)}\right] .
\end{equation}
Substituting the above expressions (\ref{g0ap}), (\ref{g1}),
(\ref{g2g0})
into Eq.\ (\ref{Gfi0}), we obtain%
\begin{equation}\label{Gfi11}
\left[ G_{1,1}^{(0)}\right] ^{-1}\approx
\frac{2J_2^{xy}}{1-\sqrt{\frac{z-2}{ z+2}}} \left\{ 1+\kappa
^2\left[\frac{\alpha \left( 2\alpha ^2-4\Delta _1\alpha +\Delta
_1^2\right) }{\Delta _1^2(2\alpha -\Delta _1)} -\frac{\left( \alpha
-\Delta _1\right) ^3}{4\Delta _1^4(2\alpha -\Delta
_1)}\right]\right\} ,
\end{equation}
and
\begin{eqnarray}
\label{Gfi12}G_{1,2}^{(0)}=G_{2,1}^{(0)} & \approx & \frac \kappa {2J_2^{xy}}%
\frac \alpha {\Delta _1^2},\\
\label{Gfi22}G_{2,2}^{(0)}(\pi ,z_b(\pi )) & \approx & \frac \alpha
{J_2^{xy}\left(\Delta _1-\alpha \right)}.
\end{eqnarray}
In the neighborhood of the point $k=\pi $, $z=z_b(\pi )$ Eq.\ (\ref
{epole}) 
may be rewritten as%
$$
-J_1^z+\left[ G_{1,1}^{(0)}\right] ^{-1}-\frac{\left[
G_{1,2}^{(0)}\right] ^2J_2^z\left(J_1^z\right)
^2}{1-J_2^zG_{2,2}^{(0)}(\pi ,z_b(\pi ))}=0.
$$
The substitution of Eqs.\ (\ref{Gfi11}), (\ref{Gfi12}) and (\ref{Gfi22})
into this equation allows to solve it with respect to $z$ and to
obtain finally Eq.\ (\ref {zbk}).

\end{document}